\documentclass[prd,aps,preprint,tightenlines,nofootinbib,superscriptaddress]{revtex4}
\usepackage{epsfig}
\usepackage{amsmath}
\input{epsf}
\usepackage{psfrag}

\usepackage[usenames,dvipsnames]{color}

\begin{document}


\vspace*{2cm}
\title{The spin dependent odderon in the diquark model}

\author{Lech Szymanowski}
\affiliation{ National Centre for Nuclear Research (NCBJ), Warsaw, Poland}

\author{Jian~Zhou}
\affiliation{\normalsize\it School of physics, $\&$ Key Laboratory of Particle Physics and Particle
Irradiation (MOE), Shandong University, Jinan, Shandong 250100, China} \affiliation{\normalsize\it
 Nikhef and Department of Physics and Astronomy, VU University Amsterdam,  De Boelelaan 1081, NL-1081 HV Amsterdam, The Netherlands}

\begin{abstract}
In this short note, we report a di-quark model calculation for the spin dependent odderon and
demonstrate that the asymmetrical color source distribution in the transverse plane of a
transversely polarized hadron plays an essential role in yielding the spin dependent odderon. This
calculation confirms the earlier finding that the spin dependent odderon is closely related to the
parton orbital angular momentum.
\end{abstract}

\maketitle

The large size of the observed transverse single spin asymmetries(SSAs) in high energy scattering
experiments~\cite{Bunce:1976yb} has stimulated a lot of theoretical developments as it allows us to
address some most fundamental aspects of QCD and gain more insight into the hadron structure as
well. The various mechanisms beyond the naive parton model~\cite{Kane:1978nd} have been proposed to
explain the large
SSAs~\cite{Sivers:1990fh,Collins:1992kk,Efremov:1981sh,Qiu:1991pp,Brodsky:2002cx,Collins:2002kn}.
The common feature of these mechanisms is that an imaginary phase required for the non-vanishing
SSAs is generated by taking into account an additional gluon exchange between the active parton and
the remanent part of the transversely polarized hadron. For a small angle scattering in the high
energy limit, the spin independent cross section is dominated by the two t-channel gluon exchange
at leading order, which can be viewed as a pomeron exchange. The three gluon exchange responsible
for the spin dependent cross section is then naturally re-interpreted as an odderon
exchange~\cite{Ryskin:1987ya,Buttimore:1998rj,Leader:1999ua,Ahmedov:1999ne}, which is a C-odd
object.  The contributions to SSAs from a such tri-gluon exchange have also been extensively
studied in the context of the collinear twist-3
framework~\cite{Ji:1992eu,Beppu:2010qn,Kang:2008qh,Kang:2008ih,Koike:2011mb,Schafer:2013opa}.

Apart from the conventional perturbative QCD
description~\cite{Bartels:1980pe,Engel:1997cga,Bartels:1999yt,Hagler:2002nf}, an odderon exchange
also can  be formulated in the dipole approach~\cite{Kovchegov:2003dm} and in the Color Glass
Condensate(CGC) framework~\cite{Hatta:2005as,Jeon:2004rk}. According to the saturation model
calculation~\cite{Kovchegov:2012ga}, the odderon exchange is absent when an unpolarized target has
uniform color source(valence quark) distribution. This observation motivated one of us to introduce
the spin dependent odderon~\cite{Zhou:2013gsa} by noticing the fact that the valence quark
distribution is strongly distorted in the transverse plane of the transversely polarized
target~\cite{Burkardt:2000za}.  In a more recent work~\cite{Boer:2015pni}, it has been found that
such spin dependent odderon is the only source of SSAs at small $x$, as the three dipole type T-odd
gluon TMD in a transversely polarized target dominating small $x$ dynamics are determined by the
spin dependent odderon. We also note that the same subject has been studied in an earlier
literature~\cite{Zakharov:1989bh}.

 The objective of this short note is to demonstrate the relation between the spin dependent
odderon and the distorted impact dependent parton distribution in a more transparent way.  For this
aim, we compute the SSA for a quark scattering off an onium that consists of one quark and one
scalar di-quark.
 As we focus on the high energy limit, we formulate  our calculation in a standard $k_T$
 factorization approach~\cite{Catani:1990eg,Collins:1991ty}(also often referred to as the high
 energy factorization), in which
 the cross section can be presented in terms of an impact factor involving the convolution
 with the wave function of the incoming hadron.
  In this work, we use the Brodsky-Hwang-Ma-Schmidt(BHMS)
 di-quark model~\cite{Brodsky:2002cx,Brodsky:2000ii} to describe the transversely polarized projectile.
 Since the BHMS model evidently incorporates the parton
orbital angular momentum effect that leads to an asymmetric impact dependent parton distribution
inside a transversely polarized hadron~\cite{Burkardt:2000za}, we use it to determine the light
cone wave function of the onium. The current work can be viewed as the one more effort
 to address the topical issue: the interplay of spin physics and saturation physics~\cite{Boer:2006rj}.

We start by fixing the relevant kinematical variables for the process under consideration.
 In the di-quark model, the partonic subprocess is expressed as the following,
\begin{eqnarray}
N(P,S_\bot)\;+\;Q(p) \rightarrow q(k)\;+\;S(r)\;+\;Q(p')
\label{process}
\end{eqnarray}
where the polarized nucleon $N$ is described within the quark-scalar diquark BHMS model, see the
Chap.~4 of Ref.~\cite{Brodsky:2000ii}. $Q$, $q$ and $S$ represent the incoming quark target, the
produced quark and the scalar di-quark, respectively.

We parameterize momenta of particles using Sudakov's light-cone vectors $p_1$ and $p_2$
($p_1^2=0=p_2^2$, $2p_1\cdot p_2=s$). The momenta of incoming nucleon $P$ with mass $M$ and
transverse polarization vector $S_\bot$, the incoming massless quark target $p$, the produced quark
$k$ with mass $m_q$ and  the helicity $\lambda_k=\pm$ and the scalar diquark $r$ with mass $m_s$
have the forms
\begin{eqnarray}
&&P=p_1+\frac{M^2}{s}p_2,\;\;\;\;\;\;p=p_2\,, \\
&&
k=z\,p_1 -\frac{(zl_\bot + v_\bot)^2-m_q^2}{sz}\,p_2+zl_\bot +v_\bot\,,\;\;\;\;
r=\bar z\,p_1 -\frac{(\bar zl_\bot - v_\bot)^2-m_s^2}{s\bar z}\,p_2+\bar zl_\bot -v_\bot\,,
\nonumber
\label{kin}
\end{eqnarray}
where $v_\bot$ is the  relative transverse momentum between the produced quark and diquark, $\bar z
=1-z$ and the momentum transfer in the $t$-channel $\Delta= p - p'$ is in high-energy kinematics
mostly the transverse vector $l_\bot$, with suppressed like ${\cal O}(1/s)$ component along $p_2$
Sudakov vector.

 It is instructive to first review how to compute the spin independent cross section of
the onium-quark scattering with a single gluon exchange as shown in Fig. 1. Using the Feynman rules
given in the Appendix I and applying the eikonal approximation to both quark and diquark lines, it
is straightforward to derive the ${\cal T}^{(1)}$ matrix for the scattering process with single
gluon exchange in momentum space, corresponding to the diagrams in the Fig.~\ref{figT1},
\begin{figure}[t]
\begin{center}
\includegraphics[width=15cm]{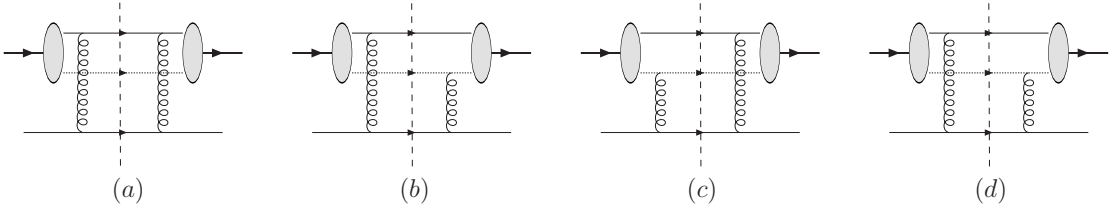}
\caption[] {Pomeron exchange between quark and onium. The solid line represent quark, while the
doted  line stands for the scalar di-quark.} \label{pomeron}
\end{center}
\end{figure}
\begin{eqnarray}
&&{\cal T}^{(1)}=2 s \lambda_s g^2 t^a_{c_k\,c_r} t^a_{c_{p'} \,c_{p}}
 \\
&&\left [ \frac{z(1-z) \bar u(z,zl_\perp+v_\perp,\lambda_k) u(P,S_\bot)}{ (zl_\perp+v_\perp)^2 -\tilde M^2} -\frac{z(1-z) \bar u(z,v_\perp-(1-z)l_\perp,\lambda_k) u(P,S_\bot)}{ (v_\perp-(1-z)l_\perp)^2 -\tilde M^2}\right ]\,\frac{\delta_{\lambda_p\,\lambda_{p'}}}{l_\bot^2}\,,
\nonumber
\end{eqnarray}
where $\lambda_p$ and $\lambda_{p'}$ are helicities of the scattered quark. Note that $\tilde M^2=
\bar z m_q^2 + z m_s^2-z\bar z M^2$ is positive. For simplicity of notation, we show as an argument
of quark spinor $\bar u(k,\lambda_k)$ only the momentum component along Sudakov vector $p_1$ and
the transverse component.  The ${\cal T}^{(1)}$ scattering amplitude expressed as a convolution in
the impact parameter $x_\bot$ (conjugate to the transverse momentum $v_\bot$)
 is given by,
\begin{eqnarray}
{\cal T}^{(1)}=2 s  g^2t^a_{c_k\,c_r} t^a_{c_{p'} c_p}    \int d^2 x_\perp \Psi(x_\perp,z) e^{ix_\perp \cdot(zl_\perp+v_\perp)}
 \left ( 1-e^{-ix_\perp \cdot l_\perp} \right )\frac{\delta_{\lambda_p\,\lambda_{p'}}}{l_\perp^2}\;,
 \label{T1}
\end{eqnarray}
with
\begin{eqnarray}
\Psi(x_\perp,z)=\lambda_s \int \frac{d^2 v_\perp}{(2\pi)^2} e^{-ix_\perp \cdot v_\perp} \frac{z(1-z)
\bar u(z,v_\perp,\lambda_k)u(P,S_\bot)}{v_\perp^2-\tilde M^2}\;
\label{wf}
\end{eqnarray}
\begin{figure}[t]
\begin{center}
\includegraphics[width=6cm]{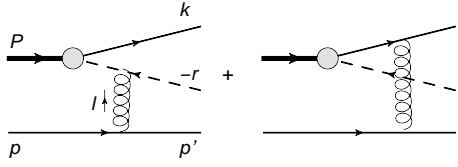}\label{figT1}
\caption[] {Feynman diagrams contributing to ${\cal T}^{(1)}$ .} \label{figT1}
\end{center}
\end{figure}
which is the nucleon wave function in impact (coordinate) representation. We proceed to compute the
product of spinors in Eq.~(\ref{wf})  using the spinor technique of Ref.~\cite{Kleiss:1986ct}. The
result expressed in terms of two orthogonal vectors of a basis in the transverse plane,
\begin{eqnarray}
\tilde e_1^\mu = S_\bot^\mu \;,\;\;\;\;\;\;\; \tilde e_2^\mu=\,\epsilon_\bot^{S_\bot \mu }
\;,\;\;\;\;\;\;\;\;\tilde e_1^2=-1=\tilde e_2^2\,,\;\;\;\;\;\tilde e_1\cdot \tilde e_2=0\;,
\label{basistilde}
\end{eqnarray}
takes the form
\begin{eqnarray}
&&\bar u(z,v_\bot,\lambda_k )u(P,\lambda_P,S_\bot) =
  \delta_{\lambda_k\,+}\frac{1}{\sqrt{2z}} \left[   Mz+m_q -\left(  \tilde e_1 + i \tilde e_2  \right)\cdot v_\bot   \right]
\nonumber \\
&&+ \delta_{\lambda_k\,-}\frac{1}{\sqrt{2z}} \left[  Mz+m_q   +\left(  \tilde e_1-i \tilde e_2
\right)\cdot v_\bot       \right]\;,
\label{wfhtilde}
\end{eqnarray}
where $\epsilon_\perp^{\nu \mu }=\,\frac{2}{s}\,\epsilon^{p_1p_2\nu \mu }$. After carrying out the
integration over $v_\bot$, one can rewrite the wave function Eq.~(\ref{wf}) as,
\begin{eqnarray}
&&\Psi(x_\perp,z)= \delta_{\lambda_k\,+}\frac{(-\lambda_s \sqrt{z}\bar z)}{2\pi \sqrt{2}} \left[
\left(Mz+m_q\right) K_0(\tilde M |x_\perp|) - \left(  \tilde e_1 + i \tilde e_2 \right)\cdot
x_\bot\,\frac{i\tilde M}{|x_\perp|} K_1(\tilde M |x_\perp|)   \right]
\nonumber \\
&&+ \delta_{\lambda_k\,-}\frac{(-\lambda_s \sqrt{z} \bar z)}{2\pi\sqrt{2}} \left[ \left( Mz+m_q
\right) K_0(\tilde M |x_\perp|) +\left(  \tilde e_1-i \tilde e_2  \right)\cdot x_\bot \,
\frac{i\tilde M}{|x_\perp|} K_1(\tilde M |x_\perp|) \right]\;, \label{wfhxtilde}
\end{eqnarray}
where  $|x_\perp|=\sqrt{-x_\bot^2}$.  This leads to the following expression for the wave function
squared,
\begin{eqnarray}
&& \sum\limits_{\lambda_k=\pm}|\Psi(x_\bot,z)|^2=
 \lambda_s^2\frac{z \bar z^2}{(2\pi)^2}\left[  \left( Mz+m_q \right)^2K_0^2(\tilde M |x_\bot|) +\tilde M^2 K_1^2(\tilde M |x_\bot|)
 \right. \nonumber \\
&& \left. \ \   + 2\left( Mz+m_q  \right) \tilde e_2\cdot x_\bot    K_0(\tilde M |x_\bot|)
  \frac{\tilde M}{|x_\bot|}K_1(\tilde M|x_\bot|) \right]\;.
\label{dcstilde}
\end{eqnarray}
With the above results one can compute the cross section
 which reads,
\begin{eqnarray}
&&
\frac{d\sigma}{d^2l_\bot}\,=\,\frac{C_F}{2\pi}\,\frac{\alpha_s^2}{(l_\bot^2)^2}\,\int\,\frac{dz}{z\bar z}\,d^2x_\bot\,\sum\limits_{\lambda_k=\pm}|\Psi(x_\bot,z)|^2(1-e^{-ix_\bot l_\bot}) (1-e^{ix_\bot l_\bot})\;.
\label{crosssec}
\end{eqnarray}
 If one integrates over the azimuthal angle of $x_\perp$ in Eq.~(\ref{crosssec}), the spin
dependent term in Eq.~(\ref{dcstilde}) drops out.
 Therefore, to derive the spin dependent cross section, one has to take into account
one additional gluon exchange as shown in Fig. \ref{odderon}.
\begin{figure}[t]
\begin{center}
\includegraphics[width=15cm]{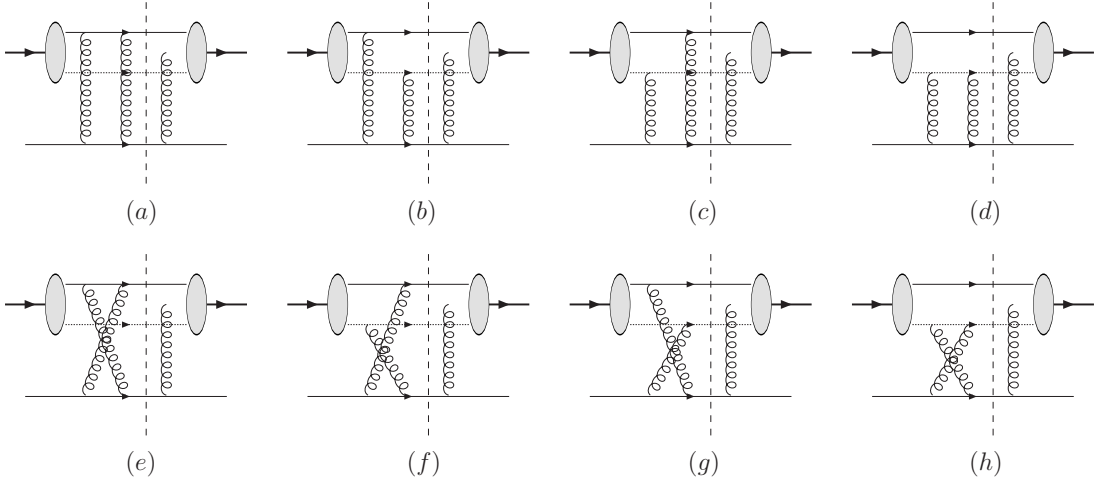}
\caption[] {Odderon exchange between quark and onium. The conjugate diagrams are not shown here.}
\label{odderon}
\end{center}
\end{figure}
Following the similar procedure as in calculation of   ${\cal T}^{(1)}$, we compute the ${\cal
T}^{(2)}$ scattering amplitude for the two gluon exchange in the symmetric $8_S$ colour octet
state. In the $k_T-$factorization approach, the ${\cal T}^{(2)}$ amplitude is expressed as a
factorized convolution in transverse momenta of $t-$channel gluons
\begin{eqnarray}
&&{\cal T}^{(2)} = \frac{1}{i\,2!}\,\left(\frac{2}{s}\right)^2\,\frac{s}{2}\,\int\,\frac{d^2l_{1\bot}}{(2\pi)^4}\frac{(-i)^2}{l_{1\bot }^2(l_\bot -l_{1\bot})^2}\,
\nonumber \\
&&\left[  \int\,d\beta_1\,{\cal S}^{(2)}(P\rightarrow q\,s)_{\mu \nu}p_2^\mu p_2^\nu \right]\,\left[\int\,d\alpha_1\, {\cal S}^{(2)}(p\rightarrow p')_{\mu' \nu'}p_1^{\mu'}p_1^{\nu'}   \right]\,.
\end{eqnarray}
It involves ${\cal S}^{(2)}(P\rightarrow q\,s)$-matrix for transition of a proton into quark and
diquark with two gluons in $8_S$ state, having longitudinal  polarizations $\sim p_2^\mu$, which is
integrated over $\beta_1$ Sudakov component of $l_1$ momentum. Similarly, ${\cal
S}^{(2)}(p\rightarrow p')_{\mu' \nu'}p_1^{\mu'}p_1^{\nu'} $ is the ${\cal S}-$matrix for transition
of target quark $p$ into outgoing quark $p'$. It is in turn integrated over $\alpha_1$ Sudakov
component of $l_1$ momentum. The combinatorial factor $1/(2!)$ assures that the ${\cal
S}^{(2)}(P\rightarrow q\,s)$-matrix element is represented as a sum of six Feynman diagrams shown
in Fig.~\ref{figup}. The expression for ${\cal T}^{(2)}$ has the form,
\begin{eqnarray}
&&{\cal T}^{(2)}= -\,\frac{i s g^4  (N_c^2-4)  t^a_{c_k c_r}  t^{a}_{c_{p'}\,c_p}\delta_{\lambda_p
\lambda_{p'}}}{2^4\,\pi^2 \,N_c}
 \\
&& \int \frac{d^2l_{1\bot} }{l_{1\bot}^2(l_\bot - l_{1\bot})^2} \,d^2x_\bot\, \Psi(x_\bot , z)
\,e^{ix_\bot (v_\bot + z l_\bot)} ( 1- e^{-i x_\bot l_{1\bot}}   )(1 - e^{ix_\bot (l_{1\bot} -
l_\bot)})\,.
 \nonumber
\label{T2coord}
\end{eqnarray}
The corresponding cross section reads,
\begin{eqnarray}
&&\frac{d\sigma}{d^2l_\bot}\,=\,\frac{i\,\alpha_s^3 C_F\,(N_c^2-4)}{2^3 \pi^2 N_c}\,
\\
&& \int\,\frac{dz\,d^2x_\bot }{z\bar z}\,\,\sum\limits_{\lambda_k=\pm}
\frac{|\Psi(x_\bot,z)|^2\;d^2l_{1\bot}}{l_{1\bot}^2 (l_\bot - l_{1\bot })^2 l_\bot^2}\,
( 1 - e^{-ix_\bot l_\bot}) ( 1- e^{ix_\bot l_{1\bot}} ) (  1 - e^{-ix_\bot (l_{1\bot} - l_\bot)}  )  + c.c.\;,
\nonumber
\end{eqnarray}
\begin{figure}[t]
\begin{center}
\includegraphics[width=11cm]{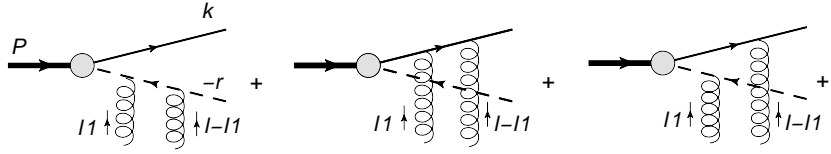}
\caption[] {Three Feynman diagrams with two gluons contributing to the transition vertex  of proton
into quark and diquark. The remaining 3 diagrams are obtained by the interchange of gluonic lines
$l_1\,\leftrightarrow\,l-l_1$ in the above diagrams.} \label{figup}
\end{center}
\end{figure}
with $|\Psi(x_\bot,z)|^2$ given by Eq.~(\ref{dcstilde}). We proceed to integrate out $l_{1\perp}$ by
the Feynman parameter method and obtain
 the expression($\bar a =1-a$),
 \begin{eqnarray}
&& \int \frac{d^2l_{1\perp}}{(2 \pi)^2}\frac{1}{l_{1\perp}^2 }\frac{1}{(l_\perp-l_{1\perp})^2 }
\left (1-e^{il_{1\perp} \cdot x_\perp}\right ) \left (1-e^{i(l_\perp -l_{1\perp})\cdot
x_\perp}\right )
\\
&=& -\frac{1}{4\pi l_\bot^2}\int_0^1 \frac{da}{a\bar a} \left \{
1+e^{il_\perp \cdot x_\perp} -\sqrt{a\bar a x_\perp^2  l_\bot^2} K_1(\sqrt{a\bar a x_\perp^2l_\perp^2} \
) \left( e^{ial_\perp \cdot x_\perp} +e^{i\bar a l_\perp \cdot x_\perp}\right ) \right \}\,, \nonumber
\end{eqnarray}
where $K_1$ is the modified Bessel function of the second kind. For a very small onium $x_\perp <<
1/l_\perp$, one can  make the following
approximation,
\begin{eqnarray}
\sqrt{a \bar a x_\perp^2l_\perp^2}  K_1(\sqrt{a \bar a x_\perp^2l_\perp^2} \ ) \approx 1\;,
\end{eqnarray}
and to perform integration over Feynman parameter $a$
\begin{eqnarray}
\hspace*{-0.4cm}\int_0^1 \frac{da}{a\bar a} \left \{
1+e^{il_\perp \cdot x_\perp} -
 e^{ial_\perp \cdot x_\perp} -e^{i\bar a l_\perp \cdot x_\perp} \right \}
 \approx \frac{(x_\bot\cdot l_\bot)^2}{2}\,(1+e^{ix_\bot\cdot l_\bot})
\,, \label{aint}
\end{eqnarray}
The cross section is simplified as,
\begin{eqnarray}
&&\frac{d  \sigma}{d^2l_\bot}\, = \frac{\,\alpha_s^3 C_F\,(N_c^2-4)}{2^3 \pi N_c}\,
 \int\,\frac{dz\,d^2x_\bot }{z\bar z}\,\,\sum\limits_{\lambda_k=\pm}
  |\Psi(x_\bot,z)|^2\frac{2(x_\bot \cdot l_\bot)^2}{  l_\bot^4}\,\sin(x_\bot\cdot l_\bot)\;.
\label{cssin}
\end{eqnarray}
 If $|\Psi(x_\perp,z)|^2$ were an azimuthal symmetric wave function,
the integration over the angle of $x_\perp$ would lead to an vanishing  cross section. However, it
has been realized long time ago that the parton distribution in the transverse plane of a
transversely polarized target is strongly distorted~\cite{Burkardt:2000za}. It can be clearly seen
from the expressions Eq.~(\ref{dcstilde}) that this is indeed the case in the quark-scalar diquark
model of a nucleon. The term in Eq.~(\ref{dcstilde}) involving $\tilde e_2\cdot x_\bot$ gives after
angular integration in Eq.~(\ref{cssin})
\begin{eqnarray}
\hspace*{-0.2cm}
\int\limits_0^{2\pi} d\phi_{x_\bot}\,\tilde e_2\cdot x_\bot (x_\bot \cdot l_\bot)^2\sin(x_\bot
\cdot l_\bot) = \tilde e_2\cdot l_\bot \,2\pi \left [J_3(|x_\bot||l_\bot|)- \frac{3
J_2(|x_\bot||l_\bot|)}{|x_\bot||l_\bot|} \right ] |x_\bot|^3|l_\bot| \,.
\end{eqnarray}
With the help of this formula, one obtains the spin dependent cross section,
\begin{eqnarray}
&&\frac{d \Delta \sigma}{d^2l_\bot}\,=  -\tilde e_2\cdot l_\bot \frac{\, \lambda_s^2 \alpha_s^3
C_F\,(N_c^2-4)}{(2\pi)^2 |l_\bot|^2 N_c} \int_0^1 dz
 \frac{\bar z  \left( Mz+m_q  \right)}{4\tilde M^4} \;,
\end{eqnarray}
where we ignored the terms suppressed by the power of $|l_\perp|/M$. This approximation is
justified for a small onium.

On the other hand, the spin dependent cross section for the quark initiated jet production in the
backward region reads~\cite{Zhou:2013gsa},
\begin{eqnarray}
\frac{d^2 \Delta \sigma}{d^2 l_\perp} =F_{x_g}(l_\perp^2)+\frac{1}{M}\epsilon_\perp^{ij}
l_{\perp}^j S_{\perp}^i O_{1T,x_g}^\perp(l_\perp^2)\;.
\end{eqnarray}
Here $F_{x_g}$ is the usual unintegrated gluon distribution, while $O_{1T}^\perp$ is the spin
dependent odderon introduced in Ref.~\cite{Zhou:2013gsa}. One thus can extract,
\begin{eqnarray}
 O_{1T,x_g}^\perp(l_\perp^2)
= -\frac{\, \lambda_s^2 \alpha_s^3 C_F\,(N_c^2-4)}{(2\pi)^2 |l_\bot|^2 N_c} \int_0^1 dz
 \frac{\bar z M \left( Mz+m_q  \right)}{4\tilde M^4}\;,
\end{eqnarray}
which is the main result of our short note. We refrain from performing the integration upon $z$ as
it is sufficient to clearly demonstrate the relation between the spin dependent odderon and the
asymmetric color source distribution in the transverse plane of the polarized target at this step.
Before summarizing the paper, few comments are in order. Though it is clear that the existence of
the spin dependent odderon relies on the polarization dependent part of the wave function which is
essentially the GPD E, one should note that  the exact relations between the odderon and the GPD E
are different in the MV model \cite{Zhou:2013gsa} and the di-quark model. In general, such
relations are model dependent. Thus, it would be very interesting to work out a model independent
relation between SSAs and  the GPD E in the future.

 At this point, we would like to comment on the phenomenogical implications of our work. Due
to the C-odd nature of the odderon exchange, it doesn't contribution to the scattering between the
onium and gluon. This implies that the SSAs caused by the spin dependent odderon disappear at
mid-rapidity which is dominated by gluon. Therefore, we anticipate that the size of SSAs rises in
the backward region of the transversely polarized target.  This observation seems to be consistent
with the measurement performed at RHIC~\cite{Bland:2013pkt}.

To summarize, we calculate the SSA in the onium-quark scattering by taking into account an odderon
exchange. It is shown that the asymmetric impact parameter dependent parton distribution inside a
transversely polarized hadron computed from the diquark model is critical for having a
non-vanishing odderon exchange. This calculation confirms that such a spin dependent odderon gives
the potential access to the parton orbital angular momentum. This is also in agreement with the
earlier observations that SSAs phenomenology is closely related to the parton orbital angular
momentum~\cite{Boros:1993ps,Burkardt:2002ks}.

\

\noindent {\bf Acknowledgments:} J. Zhou thanks Y. V. Kovchegov for stimulating discussion. J. Zhou
also thanks A. Mukherjee for helpful discussions. This research has been supported by the EU
"Ideas" program QWORK (contract 320389) and
 by grant of National Science Center, Poland, No. 2015/17/B/ST2/01838.

\begin{figure}[t]
\begin{center}
\includegraphics[width=10cm]{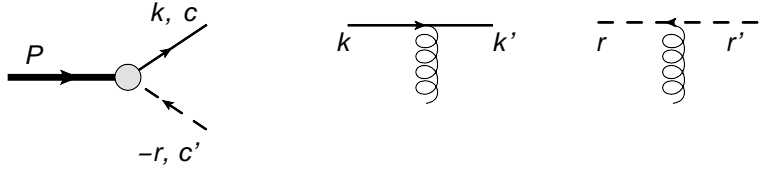}
\caption[] {Feynman rules in the diquark model.} \label{figvertices}
\end{center}
\end{figure}

\vskip.2in \noindent {\bf Appendix I:}
 \vskip.1in Based on Ref.~\cite{Brodsky:2000ii}, in the
diquark model, the interaction between the nucleon, the quark, and the scalar diquark is described
by the following Feynman rules for the nucleon-quark-diquark vertex, quark-gluon vertex,
diquark-gluon vertex,
 and the diquark and quark propagators in Fig. \ref{figvertices}, respectively,
\begin{eqnarray}
&& i \lambda_s \bar u(k,\lambda_k)u(P,S_\bot)\delta^{cc'}, \ \ \,-igt^a\gamma^\mu,\ \ \ -igt^a(r+r')^\mu,
\nonumber \\
&& \;\;\;\;\frac{i}{r^2-m_s^2+i\epsilon},\ \ \ \frac{i(\hat k+m_q)}{k^2-m_q^2+i\epsilon} \ ,
\end{eqnarray}
and the standard scalar diquark propagator, quark propagator and gluon propagator in the Feynman
gauge,
\begin{equation}
\;\;\;\;\frac{i}{r^2-m_s^2+i\epsilon},\ \ \ \frac{i(\hat k+m_q)}{k^2-m_q^2+i\epsilon}\;,\;\;\;\;\frac{-i g^{\mu \nu}\delta^{c\,c'}}{k^2+i\epsilon}\;,
\label{prop}
\end{equation}
where the subscripts $c$ and $c'$ are color indices in the adjoint representation and $\hat k=k_\mu \gamma^\mu$, $t^a$ are  $SU(N)$ gauge group generators in the fundamental representation.

\vskip.2in \noindent {\bf Appendix II:}

\vskip.1in

In this appendix, we present an alternative way of determining the wave function in the diquark
model. It is well known that the impact parameter dependent parton distribution can be
parameterized as~\cite{Burkardt:2000za},
\begin{eqnarray}
f_{q}(z,b_{\perp,q})&=&{\cal H}_q(z,b_{\perp,q}^{2})+\frac{1}{M} \epsilon_\perp^{ij} b_{\perp,q}^i
S_{\perp}^j \frac{\partial {\cal E}_q(z, b_{\perp,q}^2)}{\partial b_{\perp,q}^{ 2}}\,,
\\
f_{s}(z,b_{\perp,s})&=&{\cal H}_s(z,b_{\perp,s}^2)+\frac{1}{M} \epsilon_\perp^{ij} b_{\perp,s}^i
S_{\perp}^j \frac{\partial {\cal E}_s(z, b_{\perp,s}^2)}{\partial b_{\perp,s}^2}\,,
\end{eqnarray}
where $\cal H$ and $\cal E$ are the Fourier transform of the normal GPD $H$ and $E$. The subscript
$q$ and $s$ indicate the quark and the scalar diquark respectively.
 Using the relation,
\begin{eqnarray}
b_{\perp,q}-b_{\perp,s}=x_\perp \ \,, \ \ zb_{\perp,q}+(1-z)b_{\perp,s}=0\;.
\end{eqnarray}
and the fact that  light cone wave function and GDPs are normalized in the different way, one has,
\begin{eqnarray}
&&\int \frac{d^2x_\perp}{2\pi} \frac{dz}{2z(1-z)} |\Psi(x_\perp,z)|^2 \nonumber \\
 &&=\int d^2
[(1-z)x_\perp] dz \left \{ {\cal H}_q(z,(1-z)^2x_\perp^2)+\frac{\epsilon_\perp^{ij} x_{\perp}^i
S_{\perp}^j}{M(1-z)} \frac{\partial {\cal E}_q(z, (1-z)^2x_\perp^2)}{\partial x_\perp^2} \right
\}\;, \label{gpd}
\end{eqnarray}
where $\frac{d^2x_\perp}{2\pi} \frac{dz}{2z(1-z)} $  is the two particle phase space factor. The
expression for the GPD E in momentum space derived in the diquark model has been given in
Ref.~\cite{Meissner:2007rx}. By taking Fourier transform, one obtains,
\begin{eqnarray}
{\cal E}_q(z, (1-z)^2x_\perp^2)=\frac{\lambda_s^2}{(2\pi)^3} (m_q+zM)M K_0^2(\tilde M|x_\perp|)\;.
\end{eqnarray}
This leads to the spin dependent part of the wave function squared,
\begin{eqnarray}
\Delta |\Psi(x_\perp,z)|^2= -\epsilon_\perp^{ij} x_{\perp}^i S_{\perp}^j
\frac{2\lambda_s^2}{(2\pi)^2} (m_q+zM)(1-z)^2z
    \frac{\tilde M}{|x_\bot|}K_0(\tilde M |x_\bot|) K_1(\tilde M |x_\bot|)\;,
\label{gpd}
\end{eqnarray}
which is in complete agreement with Eq. (\ref{dcstilde}) from the direct calculation. The same result
is also found for the polarization independent piece of the wave function squared. The derivation
presented here is essentially based on the observation that both the wave function squared and the
GPDs in the position space have the clear probability interpretation.

\end {document}